\documentclass[pra,superscriptaddress,onecolumn,amsmath,amssymb,showkeys]{revtex4}
\usepackage{natbib}
\setcitestyle{square, comma, numbers,sort&compress, super}
\usepackage{amsfonts}
\usepackage{amssymb}
\usepackage{mathtools}
\usepackage{graphicx}
\usepackage{subfigure}
\usepackage{mathrsfs}
\usepackage{color}
\usepackage[normalem]{ulem}
\usepackage{natbib}
\usepackage{bbold}
\usepackage{placeins}
\usepackage{braket}
\usepackage{soul,xcolor}
\usepackage{epstopdf}
\usepackage{tabu}
\usepackage{array}
\usepackage{bm}
\usepackage{upgreek}
\usepackage{multirow}
\usepackage[utf8]{inputenc}
\usepackage[colorlinks = truelinkcolor = blue, urlcolor  = blue, citecolor = blue, anchorcolor = blue]{hyperref}
\hypersetup{
	colorlinks   = true, 
	urlcolor     = blue, 
	linkcolor    = blue, 
	citecolor   = blue 
}

\begin{document}
\title{Probing inequivalent forms of Legget-Garg inequality in subatomic systems}
	\author{Javid Naikoo}
\email{naikoo.1@iitj.ac.in}
\affiliation{Indian Institute of Technology Jodhpur, Jodhpur 342011, India}

\author{Swati Kumari}
\email{ swatipandey084@gmail.com}
\affiliation{National Institute of Technology Patna, Ashok Rajpath, Patna, Bihar 800005, India}

\author{Subhashish Banerjee}
\email{subhashish@iitj.ac.in}
\affiliation{Indian Institute of Technology Jodhpur, Jodhpur 342011, India}

\author{A. K. Pan}
\email{akp@nitp.ac.in}
\affiliation{National Institute of Technology Patna, Ashok Rajpath, Patna, Bihar 800005, India}

\begin{abstract}
	We study various formulations of Leggett-Garg inequality (LGI), specifically, the  Wigner and Clauser-Horne forms of LGI,  in the context of subatomic systems, in particular, three flavor neutrino as well as meson systems. The optimal forms of various LGIs for either neutrinos or mesons are seen to depend on measurement settings. For the neutrinos, some of these inequalities can be written completely in terms of  experimentally measurable probabilities. Hence, the  Wigner and Clauser-Horne forms of LGI are found to be more suitable as compared to the standard LGI from the experimental point of view for the neutrino system. Further, these inequalities exhibit maximum quantum violation  around the energies roughly corresponding to the maximum neutrino flux.  The Leggett-Garg type inequality is seen to be more suited for the meson dynamics. The meson system being inherently a decaying system, allows one to see the effect of decoherence on the extent of violation of various inequalities. Decoherence is observed to reduce the degree of violation, and hence the nonclassical nature of the system.
\end{abstract}
\keywords{Neutrino, Neutral Mesons, Leggett-Garg inequality.\\
	 Submitted to: Journal of Physics G: Nuclear and Particle Physics.}
\maketitle

\section{Introduction}\label{intro}
 Testing the validity of quantum mechanics at the macroscopic level has gained much attention over the recent years. The emergence of our everyday world view of reality  from the laws of  quantum mechanics is still a debatable issue in quantum foundations. Historically, this question was first posed by Schrondinger  through his famous cat thought experiment \cite{sch}. Several approaches have been adopted  to address this problem, for example, experimental realization of the quantum coherence of large objects \cite{arndt}, the decoherence models \cite{zur}, in which system-environment interactions reduce the degree of coherence. Other approach is the coarse-grained measurements \cite{bruk}, where by putting constraints on the measuring apparatus leads to the emergence of classicality. However, these approaches do not fundamentally answer the question  whether macrorealism, in principle, is compatible with quantum mechanics. The concept of macrorealism is based on our everyday experience of macroscopic world in which  the properties of macroscopic objects  exist irrespective of the observation. A quantitative test for investigating localrealism was devised by John Bell in the form of Bell's inequality \cite{bell64}, which is violated by quantum systems and hence nullifies the existence of localrealistic description for such systems. Motivated by the Bell's inequality, Leggett and Garg \cite{leggett85} formulated a class of inequalities based on the notions of macrorealism, which provides an elegant scheme for experimentally testing the compatibility between macrorealism and the axioms of quantum theory. 
The concept of macrorealism consists of two main assumptions \cite{leggett85} which seem reasonable in our everyday world: (a) \emph{ Macrorealism per se } (MR): If a macroscopic system has two or more macroscopically distinguishable ontic states available to it, then the system remains in one of those states at all instants of time. (b) \emph{Noninvasive measurability} (NIM): The definite ontic state of the macrosystem is determined without affecting the state itself or its possible subsequent dynamics. Consider a dichotomic observable $\hat{M}$ having outcomes $\pm1$, and measurements performed at time $t_1$,  $t_2$  and  $t_3$, which in turn can be considered as the measurement of the observables $\hat{M}_1 = \hat{M}(t_1)$, $\hat{M}_2 = \hat{M}(t_2)$, and  $\hat{M}_3 = \hat{M}(t_3)$, respectively.
A  measurement of the observables $\hat{M}_1$ , $\hat{M}_2$, and $\hat{M}_3$ must lead to definite outcomes $+1$ or $-1$ at all instants of time in accordance with  the assumption of MR. The  NIM condition, in this context, implies that the outcomes of a measurement of $\hat{M}_2$ or $\hat{M}_3$ remain unaffected due to the measurement of $\hat{M}_1$, and so on. One can then formulate the standard Leggett-Garg inequalities (LGIs) as
\begin{equation}
\label{lgi}
K_{3} = m_{1}m_{2}\langle M_{1} M_{2}\rangle+ m_{2} m_{3}\langle M_{2}M_{3} \rangle  - m_{1}m_{3}\langle M_{1}M_{3}\rangle \leq1,
\end{equation}
where $m_{1},m_{2},m_{3}=\pm1$. It is well studied that in quantum theory $(K_{3})_{Q}>0$ for a suitable choice of observables, even for a qubit system. The LGIs have been investigated in various studies both on the theoretic 
\cite{gangopadhyay2013probing,emary2013leggett,avis2010leggett,budroni2013bounding,kofler2007classical,MesonPRD,LGtIprd} as well as experimental \cite{dressel2011experimental,goggin2011violation,lambert2010distinguishing,tittel1998experimental,formaggio2016violation,DayaBay} fronts.  It is well known that neutrino oscillations can exhibit coherence over large distances owing to their weakly interacting nature \cite{formaggio2016violation}. This makes them promising future candidates for carrying out quantum information tasks.  Therefore, analyzing the nonclassical properties in this system, in terms of experimentally verifiable measures, is important both from theoretic as well as application point of view. Further, the study of nonclassical measures like LGIs can reveal important information about the underlying dynamics in decaying systems like neutral mesons \cite{MesonPRD}. This motivates us to study various avatars of LGI which are amenable to experimental verification and at the same time show  prominent  violations within the experimental parameters considered in this work.

The plan of the paper is as follows: We briefly discuss some variants of LGI and revisit the dynamics of neutrino and meson systems, relevant to our work. This is followed by a study of these different  forms of LGIs on these systems. We finally make our conclusions.

\section{Various avatars of LGI and their experimental relevance}

\subsection{Variants of LGI}
In recent times, various other formulations of LGIs, viz., Entropic LGI \cite{UshaELGI,naikoo2018entropic}, Wigner \cite{saha} and Clauser-Horne \cite{swati19} form of LGIs has also been proposed. A new variant of LGIs has also been proposed providing the quantum violation upto the algebraic maximum \cite{pan18}. Note that the assumptions of macrorealism \textit{per se} and non-invasive measurability imply the existence of joint probability distribution in a macrorealist model. From the assumptions of joint probability and non-invasive measurability, we obtain the pairwise statistics of measurement of $\hat{M_2}$ and  $\hat{M_3}$ having outcome $m_{2}$ and $m_{3}$ as 
$P(m_{2},m_{3})=\sum_{m_{1}=\pm}P(m_{1},m_{2},m_{3})$ and similarly for others. We can write the expression, 
\( P(-m_{1},m_{2})+P(m_{1},m_{3})-P(m_{2},m_{3})=P(-m_{1},m_{2},-m_{3})+P(m_{1},-m_{2},m_{3}) \) . By invoking the non-negativity of the probability, Wigner form of LGIs can be derived as
\begin{eqnarray}
\label{w1}
P(m_{2},m_{3})-P(-m_{1},m_{2})-P(m_{1},m_{3})\leq 0.		
\end{eqnarray}
One can obtain eight variants Wigner form of LGIs from Ineq. (\ref{w1}).
Similarly, sixteen more inequalities can be derived from
\begin{eqnarray}
\label{w2}
P(m_{1},m_{3})-P(m_{1},-m_{2})-P(m_{2},m_{3})\leq 0,
\end{eqnarray}
\begin{eqnarray}
\label{w3}
P(m_{1},m_{2})-P(m_{2},-m_{3})-P(m_{1},m_{3})\leq 0.
\end{eqnarray}
Thus one has twenty four variants of Wigner form of LGI characterized by different measurement settings. This richness turns out to be very useful especially in systems where experimental constraints put limitation on arbitrary preparation and detection process, viz., in subatomic systems like neutrinos and mesons. Some of us have recently shown that Wigner form of LGIs are stronger than the standard LGIs \cite{pan17,swati17}.\par
The single marginal statistics of the measurement of the observable, for example, probability of getting outcome, when $M_{2}$ measurement is performed can be obtained as
$P(m_{2})=\sum_{m_{1},m_{3}=\pm}P(m_{1},m_{2},m_{3})$ and similarly for $P(m_{1})$ and  $P(m_{3})$. By combining single and pair-wise statistics, we can get the expression,
$P(m_{1},m_{3})+P(m_{2})-P(m_{1},m_{2})-P(m_{2},m_{3})=P(m_{1},-m_{2},m_{3})+P(-m_{1},m_{2},-m_{3})$, which gives
\begin{eqnarray}
\label{ch1}
P(m_{1},m_{2})+P(m_{2},m_{3})-P(m_{1},m_{3})-P(m_{2})\leq0.
\end{eqnarray} 
Inequality (\ref{ch1}) can lead to eight variants of Clauser-Horne form of LGIs \cite{swati19}. Similarly, sixteen  more inequalities can be derived in this manner.
In compact notation, we can write,
\begin{eqnarray}
\label{ch2}
P(m_{1},m_{3})+P(m_{1},m_{2})-P(m_{2},m_{3})-P(m_{1})\leq0,
\end{eqnarray}
\begin{eqnarray}
\label{ch3}
P(m_{1},m_{3})+P(m_{2},m_{3})-P(m_{1},m_{2})-P(m_{3})\leq0.
\end{eqnarray}
Note that in the Wigner form of LGIs only pair-wise probabilities are involved but in Clauser-Horne form of LGIs single probabilities are also involved along with pair-wise ones.
Wigner and Clauser-Horne forms of LGIs can be shown to be equivalent to standard LGIs in macrorealist model, but inequivalent in quantum theory \cite{swati19}. 
In order to show this, we write the pair-wise joint probability, for example, $P(m_{2},m_{3})$ in the moment expansion is given by
\begin{equation}
\label{i1}
P(m_{2},m_{3})=\frac{(1+ m_{2} \langle M_{2}\rangle+ m_{3}\langle M_{3}\rangle+m_{2}m_{3} \langle M_{2}M_{3}\rangle)}{4}.
\end{equation}
Similarly, the single probabilities, for example, $P(m_{3})$ can be written as 
\begin{equation}
\label{i2}
P(m_{3})=\frac{(1+ m_{3}\langle M_{3}\rangle)}{2},
\end{equation}
 where $P(m_{3})=\sum_{m_{1},m_{2}=\pm} P(m_{1},m_{2},m_{3})$.

Putting the relevant pair-wise joint probabilities (as in Eq. (\ref{i1})) into that left hand side of the  Ineqs. (\ref{w1}-\ref{w3}) and Ineqs. (\ref{ch1}-\ref{ch3}) one obtained the standard LGIs given by Ineq. (\ref{lgi}). Thus, Wigner and Clauser-Horne forms of LGIs are equivalent to standard LGIs in a macrorealisic theory.
 
Now, let us examine the equivalence among various LGIs in quantum scenario. Given a density matrix $\rho$, in quantum theory a pair-wise probability \cite{hall14} can be written as
\begin{eqnarray}
\label{p1}
P_{Q}(m_{1},m_{2})=\frac{1}{4}(1+ m_{1} \langle M_{1}\rangle_{Q}+ m_{2}\langle M_{2}^{(1)}\rangle_{Q}
+m_{1} m_{2} \langle M_{1}M_{2}\rangle_{Q}),
\end{eqnarray}
and a single probability is given by
\begin{equation}
\label{p2}
P_{Q}(m_{1})=\frac{(1+ m_{1}\langle M_{1}\rangle_{Q})}{2},
\end{equation}
 where the superscript in $\langle M_{2}^{(1)}\rangle_{Q}$ denotes that the measurement of $M_2$ in quantum theory is disturbed by the prior measurement $M_{1}$.
\\
\\
Now, corresponding to $24$ Wigner form of LGIs given by Ineqs.  (\ref{w1})-(\ref{w3})
using Eq.(\ref{p1}) and similar quantities, the left hand side in quantum theory
\begin{eqnarray}
\label{t1}
(W^{3})_{Q}=|\langle M_{2} \rangle-\langle M_{2}^{(1)}\rangle|+ |\langle M_{3}^{(2)}\rangle-\langle M_{3}^{(1)}\rangle|+(LG^{3})_{Q},
\end{eqnarray}
where $(LG^{3})_{Q}$ quantum expression of $LGI$ given by Ineq. (\ref{lgi}). If the measurement of $M_{1}$ does not disturb the statistics of $M_2$, then $\langle M_{2} \rangle=\langle M_{2}^{(1)}\rangle$ and if prior measurements do not disturb the statistics of $M_3$, so that, $\langle M_{3}^{(2)}\rangle=\langle M_{3}^{(1)}\rangle=\langle M_{3}\rangle$. In that situation, Eq.(\ref{t1}) reduces to $(LG^{3})_{Q}$ only and we can say Wigner form of LGIs are equivalent to the standards ones in quantum theory. But in quantum theory,  $|\langle M_{2} \rangle-\langle M_{2}^{(1)}\rangle|\neq0$ and $|\langle M_{3}^{(2)} \rangle-\langle M_{3}^{(1)}\rangle|\neq0$, in general. Hence, from Eq. (\ref{t1}), we can say that the violation of standard LGIs implies the violation of Wigner form of LGIs, but the converse is not true. Hence, Wigner form of LGIs are stronger than the standard LGIs and captures the notion of macrorealism better than standard LGIs. Similarly, it can be shown that the Clauser-Horne form of LGIs 
are also stronger than the standard LGIs, a detailed discussion is given in \cite{swati19}.

\subsection{Experimental relevance}

Recently, the study of LGIs and their variants has gained significant interest in the context of subatomic systems, particularly, flavor oscillations in neutrinos and mesons  \cite{LGtIneutrino,MesonPRD,LGtIprd}.  The  LGIs in the context of three flavor neutrino oscillations, cannot be expressed completely in terms of the measurable survival and transition probabilities \cite{LGtIprd}, thereby making it difficult to verify them experimentally. Same problem is encountered while dealing with neutral meson systems. One can bypass such experimental constraints by invoking the assumption of \textit{stationarity}, leading to a class of LG type inequalities (LGtIs) \cite{huelga1995proposed,huelga1996temporal,emary2013leggett,LGtIneutrino}. For stationarity to hold, the  following set of conditions must be satisfied:  (a) macrorealism, (b) time translation invariance of probabilities, i.e., $P(\psi, t + t_0 | \psi, t_0) = P(\psi,t | \psi, 0)$, where $P(x,t|y,s)$ stands for the conditional probability for a system to be in state $x$ at time $t$ given that it was in state $y$ at  time $s$, (c) the underlying dynamics is \textit{Markovian}, and (d) the system is prepared in a well defined state at time $t=0$. However, this puts constraints on the type of dynamics that could be investigated. Therefore, it would be worthwhile to look for formulations that could be expressed {\it completely} in terms of experimentally measurable quantities while allowing for all the basic axioms. It turns out that some of the variants of Wigner and CHSH inequalities, discussed above, are able to accomplish this. They can be completely expressed in terms of measurable probabilities without making any further assumptions. This sets the tone for the present work as well as brings out its relevance.

Here, we probe Wigner and Clauser-Horne forms of LGIs in the context of three flavour neutrino and meson systems.  In both the formulation of LGIs, most of the inequalities contain non-measurable terms, as in the case of the standard LGIs \cite{LGtIprd}. However, in the context of neutrino oscillations, we find that some of these inequalities can be expressed solely  in terms of the experimentally measurable quantities, i.e., neutrino survival and transition probabilities. This is a very attractive feature which should help in probing  foundational  issues in subatomic physics.  In case of neutrinos, the relevant inequalities are analyzed for ongoing experiments NO$\nu$A (NuMI Off-axis $\nu_e$ Appearance), T2K (Tokai to Kamioka) and the  upcoming experiment DUNE (Deep Underground Neutrino Experiment). These experiments have specific baseline and   energy range, adapted here.


\section{Dynamics of  neutrino and meson systems}\label{dynamics}

In this section, we briefly review the dynamics of neutrino system in the context of three flavor neutrino oscillations. We also discuss the time evolution of  the neutral meson ($K^0$). The neutrino state time evolution is unitary; however the meson system being decaying in nature is a non-unitary system and is dealt with using the approach of open quantum systems \cite{BanerjeeBook}.
\subsection{Three flavor neutrino system}
When dealing with the three flavor scenario of neutrino oscillation \cite{Pontecorvo}, one represents a general neutrino  state either in the \textit{flavor basis} $\{ \ket{\nu_\alpha}\}$ ($\alpha=e, \mu, \tau$) or in the \textit{mass basis} $\{ \ket{\nu_k}\}$ ($k=1, 2, 3$)
\begin{equation}
\ket{\Psi} = \sum\limits_{\alpha=e, \mu,  \tau} \psi_\alpha \ket{\nu_\alpha} = \sum\limits_{k= 1, 2, 3} \psi_k \ket{\nu_k}.
\end{equation}
The expansion coefficient in the two representations are connected by the so called Pontecorvo-Maki-Nakagawa-Sakata  (PMNS) matrix as follows
\begin{equation}\label{ftom}
\psi_\alpha = \sum\limits_{k=1,2,3} U_{\alpha, k}  \psi_k.
\end{equation}
Here, $U_{\alpha, k}$ are the element of the PMNS matrix. Later can be parametrized in many ways, one that is often used in the literature  \cite{chau1984comments,giunti2007fundamentals} is given below
\begin{equation}
\boldsymbol{U} = 
\begin{pmatrix}
c_{12} c_{13} & s_{12} c_{13} & s_{23} e^{-i \delta} \\ - s_{12}c_{23} - c_{12} s_{23}s_{13} e^{i\delta} & c_{12}c_{23}-s_{12}s_{23}s_{13} e^{i\delta} & s_{23}c_{13} \\ s_{13}s_{23} - c_{12}c_{23}s_{13} e^{i\delta} & -c_{12}s_{23}-s_{12}c_{23}s_{13} e^{i\delta} & c_{23}c_{13}\end{pmatrix}.
\end{equation}
Here $c_{ij} = \cos\theta_{ij}$, $s_{ij} = \sin\theta_{ij}$, and the parameters $\theta_{ij}$ and $\delta$  are the mixing angles and the $CP$ violating phase \cite{BKayser}, respectively.  In this work, the  values of  mixing angles used are  $\theta_{12} =33.4^o$,  $\theta_{13} = 8.50^o$, and $\theta_{23} = 42.30^o$ (assuming normal mass-ordering)   \cite{patrignani2016chin}. The Eq. (\ref{ftom}) can be written in matrix form as $ \boldsymbol{\psi}_{f} = \boldsymbol{U}~  \boldsymbol{\psi}_m $, with $ \boldsymbol{\psi}_{f}$ and $ \boldsymbol{\psi}_{m}$ (where $f$ and $m$ signify the flavor and mass, respectively) are the column vectors of the expansion coefficients. The massive eigenstates evolve according to the Schrodinger equation, such that $ \boldsymbol{\psi}_m(t)= \boldsymbol{E}~  \boldsymbol{\psi}_m(0)$. Here $\boldsymbol{E} = {\rm diag}.[e^{iE_1 t}, ~e^{iE_2 t}, ~ e^{iE_2 t}]$ is the diagonal matrix and $E_1$, $E_2$ and $E_3$ are the energies corresponding to the massive eigenstates $\ket{\nu_1}$, $\ket{\nu_2}$ and $\ket{\nu_3}$, respectively. One can now connect the flavor state at time $t=0$ and some later time $t$ by the following relation
\begin{equation}\label{Uf}
 \boldsymbol{\psi}_f (t) = \boldsymbol{U}~ \boldsymbol{E}~ \boldsymbol{U}^{-1}~  \boldsymbol{\psi}_f (0) = \boldsymbol{U}_f(t)   \boldsymbol{\psi}_f (0).
\end{equation}
We  call $\boldsymbol{U}_f(t)$ the flavor evolution operator, which takes a flavor state at time $t=0$ to some later time $t$. It is worth mentioning here that the above formalism is valid only for the neutrino propagation in vacuum. In order to carry out the analysis in the context of the neutrino experiments, one has take into account the matter effect as well. A detailed account on how to construct the time evolution operator in such a case, can be found in \cite{ohlsson2000three} and  references therein.  It turns out that in presence of matter effects, the operator $\bm{U}_f$, apart from the mixing angles and mass-square differences, also depends on the matter density parameter $A = \pm \sqrt{2} G_F N_e$. Here, $G_F$ is the Fermi weak coupling constant and $N_e$ is the electron density. The sign $+$ and $-$ is considered for neutrinos and antineutrinos, respectively.\par
\subsection{Neutral meson $K^0 - \bar{K}^0$ system}
In this subsection, we revisit the formalism of the operator sum representation which is an important tool used to describe the dynamics of the decaying neutral meson system. This will be followed by a discussion in the context of $K^0 - \bar{K}^0$ system.\par
\textit{Operator sum representation:} 
The time evolution of a closed system can be  describe by a unitary operator. However, this is not true for an open system and one often resorts to what is called the \textit{operator sum representation} (OSR) in terms of the Kraus operators \cite{kraus1983states}. The OSR has proved to be a powerful tool for dealing with open quantum systems  
\cite{breuer2002theory,BanerjeeBook,nielsenchuangbook,omkar2012operator,omkar2015operator,banerjee2016quantum}. The total Hilbert space is $\mathcal{H}_S \otimes \mathcal{H}_E$ with the constraint that the system and environment start in the product state at time $t=0$, that is, $\rho(0) = \rho_S \otimes \rho_E$. The time evolution of the combined system is then governed by the unitary operator $ U_{SE}(t)$ as follows
\begin{equation}
\rho(t) = U_{SE}(t) \rho(0) U_{SE}^{\dagger}(t).
\end{equation}
Usually one is interested in the dynamics of  the system of interest and  the environmental degrees of freedom are traced out
\begin{equation}
\rho_S(t) = Tr_E\{U_{SE}(t)\rho(0) U_{SE}^{\dagger}(t)\}. \label{rho_S}
\end{equation}
One may write this reduced state in the following representation
\begin{equation}
\rho_S(t) = \sum_{i} \mathcal{K}_i(t) \rho_S(0) \mathcal{K}_i^{\dagger}(t).
\end{equation}
The unitary nature of $U_{SE}(t)$ ensures that $\sum_{i} \mathcal{K}_i(t) \mathcal{K}_i^{\dagger}(t) = \mathbb{1} $, implying that the evolution of  $\rho_S(t)$  has a Kraus representation and is completely positive.

\textit{Time evolution of $K^0-\bar{K}^0$ meson system:}	
Here, we spell out the open system dynamics of the $K^0-\bar{K}^0$ system \cite{banerjee2016quantum}. The Hilbert space of the total system is given by the direct sum $\mathcal{H}_{K^0}  \oplus \mathcal{H}_{0}$ \cite{caban2005unstable,ABUDecho,Alok:2013sca} spanned by the orthonormal vectors $\ket{K^0}$, $\ket{\bar{K}^0}$ and $\ket{0}$ (denoting the vacuum state)

\begin{equation}
\ket{K^0} = \begin{pmatrix}
1  \\
0   \\
0
\end{pmatrix}; \quad \ket{\bar{K}^0} = \begin{pmatrix}
0  \\
1   \\
0 
\end{pmatrix}; \quad  \ket{0} = \begin{pmatrix}
0  \\
0   \\
1
\end{pmatrix}. \label{basis}
\end{equation} 
The states $\{\ket{K^0}, \ket{\bar{K}^0} \}$ are the eigenstates of the strangeness operator $\hat{S}$;  $\hat{S} \ket{K^0} = \ket{K^0},~ \hat{S} \ket{\bar{K}^0} = -\ket{\bar{K}^0},~\hat{S} \ket{0} = 0$. These are related to charge-parity ($CP$)  eigenstates $\{ \ket{K_1^0}, \ket{\bar{K}^0_2}\}$ as follows 
\begin{equation}\label{MassFlavorStates}
\ket{K^0_1} =  \frac{\ket{K^0} +  \ket{\bar{K}^0}}{\sqrt{2}}, \qquad \ket{K^0_2} = \frac{\ket{K^0} -  \ket{\bar{K}^0}}{\sqrt{2}}.
\end{equation} 
  Further, the $CP$ eigenstates are related to what are known as short and long lived eigenstates $\{\ket{K_S}, \ket{K_L}\}$ as follows
\begin{equation}
\ket{K_S} = \frac{1}{\sqrt{1 + |\epsilon|^2}} (\ket{K_1^o} + \epsilon \ket{\bar{K}^o_2}), \qquad \ket{K_L} = \frac{1}{\sqrt{1 + |\epsilon|^2}} (\epsilon \ket{K^o_1} +  \ket{\bar{K}^o_2}),
\end{equation}
where $\epsilon$ is a measure of the departure from perfect $CP$ invariance. 
The complete positivity demands the following  OSR  \cite{kraus1983states}

\begin{equation}
\rho(t) = \sum_{i=0} \mathcal{K}_{i}(t) \rho(0) \mathcal{K}^{\dagger}_{i}(t), \label{operator_sum_rep}
\end{equation}
where the Kraus operators have the following form \cite{MesonPRD}

\begin{align*}
\mathcal{K}_0 &= \ket{0}\bra{0},\\
\mathcal{K}_1 &= \mathcal{C}_{1+}\bigg[ \ket{K^0}\bra{K^0} + \ket{\bar{K}^0}\bra{\bar{K}^0}\bigg] + \mathcal{C}_{1-}\bigg[ \frac{1+ \epsilon}{1- \epsilon}\ket{K^0}\bra{\bar{K}^0} + \frac{1- \epsilon}{1 + \epsilon}\ket{\bar{K}^0}\bra{K^0} \bigg],\\
\mathcal{K}_2 &= \mathcal{C}_2 \bigg[ \frac{1}{1  + \epsilon} \ket{0}\bra{K^0} + \frac{1}{1  - \epsilon} \ket{0}\bra{\bar{K}^0} \bigg], \\
\mathcal{K}_3 &= \mathcal{C}_{3+} \frac{1}{1 + \epsilon}\ket{0}\bra{K^0} + \mathcal{C}_{3-} \frac{1}{1 - \epsilon}\ket{0}\bra{\bar{K}^0} , \\
\mathcal{K}_4 &= \mathcal{C}_4 \bigg[ \ket{K^0}\bra{K^0} + \ket{\bar{K}^0}\bra{\bar{K}^0} + \frac{1 + \epsilon}{1 - \epsilon}\ket{K^0}\bra{\bar{K}^0} + \frac{1 - \epsilon}{1 + \epsilon}\ket{\bar{K}^0}\bra{K^0} \bigg],\\
\mathcal{K}_5 &= \mathcal{C}_5 \bigg[ \ket{K^0}\bra{K^0} + \ket{\bar{K}^0}\bra{\bar{K}^0} - \frac{1 + \epsilon}{1 - \epsilon}\ket{K^0}\bra{\bar{K}^0} - \frac{1 - \epsilon}{1 + \epsilon}\ket{\bar{K}^0}\bra{K^0} \bigg].
\end{align*}
The coefficients appearing in the above equations are given by 
\small
\begin{align}
\begin{rcases}
\mathcal{C}_{1\pm} &= \frac{1}{2} \left[ e^{-(2 i m_S + \Gamma_S + \lambda) t/2} \pm e^{-(2 i m_L + \Gamma_L + \lambda) t/2} \right],\\
\mathcal{C}_2  &= \sqrt{\frac{1+ |\epsilon|^2}{2} \big( 1 - e^{- \Gamma_S t} - \delta_L^2  \frac{|1 - e^{-(\Gamma + \lambda - i \Delta m )t}|^2}{1 - e^{-\Gamma_L t}}}\big),\\
\mathcal{C}_{3\pm} &= \sqrt{\frac{1+ |\epsilon|^2}{2(1 - e^{-\Gamma_L t})}} \big[1 - e^{-\Gamma_L t}  \pm (1 - e^{-(\Gamma + \lambda - i \Delta m)t}) \delta_L\big],\\
\mathcal{C}_{4} &= \frac{e^{-\Gamma_S t/2}}{2} \sqrt{1 - e^{-\lambda t}},\\
\mathcal{C}_{5} &= \frac{e^{-\Gamma_L t/2}}{2} \sqrt{1 - e^{-\lambda t}}. 
\end{rcases}
\end{align}
\normalsize
Starting at time $t=0$ with state $\rho_{K^0}(0) = \ket{K^0}\bra{K^0}$ or $\rho_{\bar{K}^0}(0) = \ket{\bar{K}^0}\bra{\bar{K}^0}$, the state at some later time $t$, is given by 
\small
\begin{equation}
\rho_{K^0}(t) = \frac{1}{2}e^{-\Gamma t} \begin{pmatrix}
a_{ch} + e^{-\lambda t} a_{c}                   & (\frac{1 - \epsilon}{1 + \epsilon})^* (-a_{sh} - i e^{- \lambda t} a_s)     &       0 \\\\
(\frac{1 - \epsilon}{1 + \epsilon}) (-a_{sh} + i e^{-\lambda t} a_s)  & |\frac{1 - \epsilon}{1 + \epsilon}|^2 a_{ch} - e^{-\lambda t} a_{c}         &       0  \\\\
0                                       &                0                                      &     \rho_{33}(t)
\end{pmatrix} \label{rhoBt},
\end{equation}
and 
\begin{equation}
\rho_{\bar{K}^0}(t) = \frac{1}{2} e^{-\Gamma t} \begin{pmatrix}
|\frac{1 + \epsilon}{1 - \epsilon}|^2 (a_{ch} - e^{- \lambda t} a_c)      &   (\frac{1 + \epsilon}{1 - \epsilon}) (-a_{sh} + i e^{- \lambda t} a_{s})  &  0  \\\\
(\frac{1 + \epsilon}{1 - \epsilon})^* (-a_{sh} -i e^{- \lambda t} a_{s})  &    a_{ch} + e^{- \lambda t} a_c                      &  0   \\\\
0                                             &           0                                          &  \tilde{\rho}_{33}(t) \\
\end{pmatrix}  \label{rhoBbart}.
\end{equation}
\normalsize
Here, $a_{ch} = \cosh[{\frac{\Delta \Gamma t}{2}]}$,  $a_{sh} = \sinh{[\frac{\Delta \Gamma t}{2}]}$  and $a_{c}=\cos{[\Delta m t]}$, $a_{s} = \sin{[\Delta m t]}$ and $\epsilon$ is  the CP violating parameter. $\Delta\Gamma = \Gamma_S - \Gamma_L$ is the difference of the decay width $\Gamma_S$ (for $K^0_S$ ) and  $\Gamma_L$ (for $K^0_L$). $\Gamma = \frac{1}{2}(\Gamma_L + \Gamma_H)$ is the average decay width. The mass difference between the long and short lived states is given by $\Delta m = m_L - m_S$, where $m_L$ and $m_S$ are the masses of $K^0_L$ and $K^0_S$ states, respectively. The \textit{decoherence} parameter $\lambda$ is proportional to the strength of the interaction between the one particle system and its environment \cite{ABUDecho}. The above discussed formalism is used in the next sections to analyze Wigner and Clauser-Horne form of LGI for these systems.

\section{Quantum violation of Wigner and Clauser-Horne form of LGIs in neutrino system}\label{WLGIandCHinNeutrino}
\begin{figure}[t!]
	\centering
	\includegraphics[width=50mm]{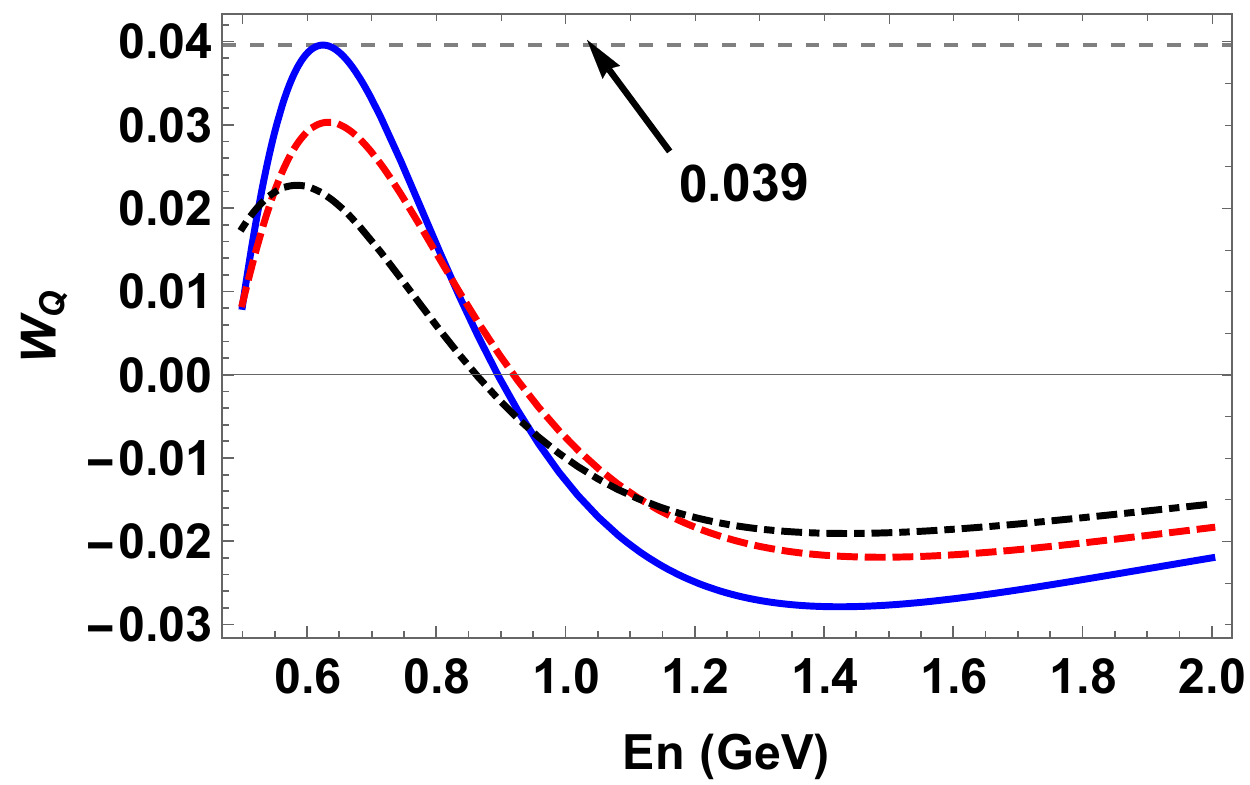}
	\includegraphics[width=50mm]{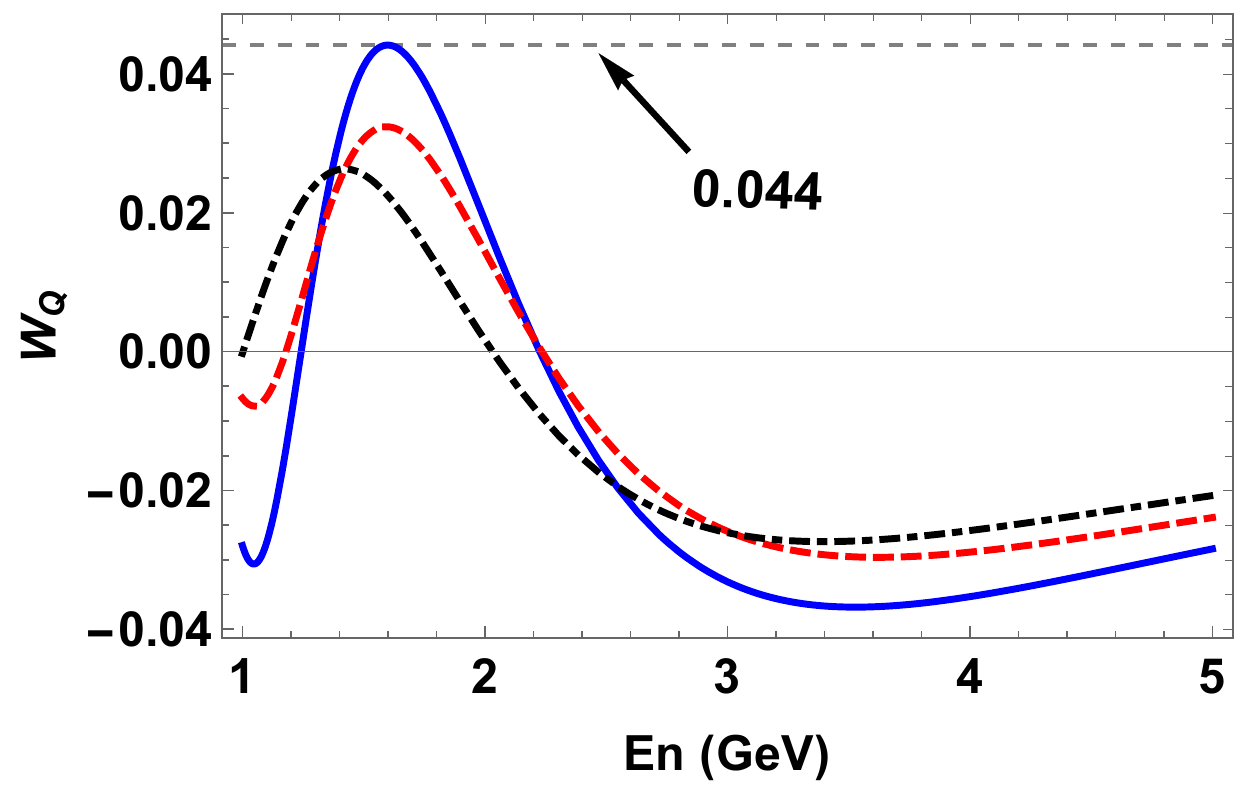}
	\includegraphics[width=50mm]{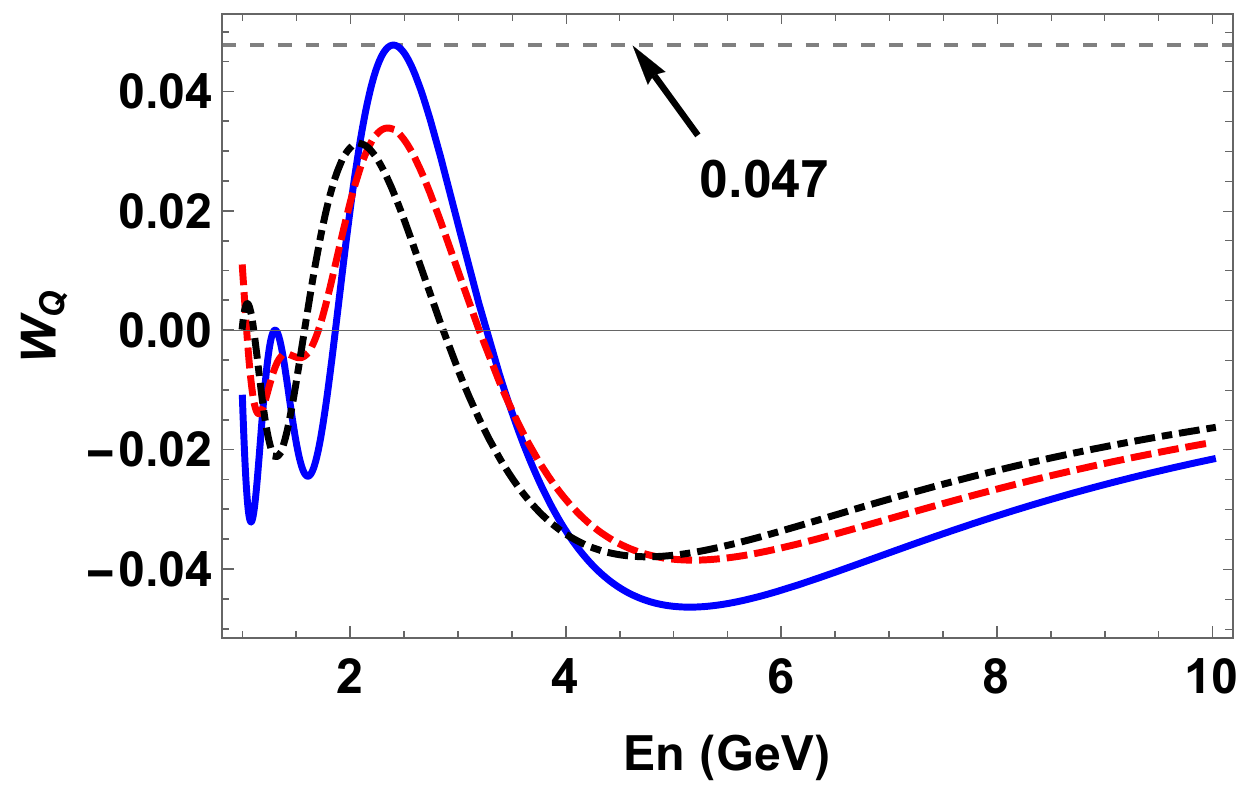}
	\caption{(color online) Wigner form of LGI (bottom panel) (Eq. (\ref{W4})) in neutrino system for different experimental set ups vz., $T2K$ (left), $NO\nu A$ (middle) and DUNE (right), plotted with respect to the neutrino energy (En) in GeV. The baseline of 295 km, 810 km and 1300 km are used for  $T2K$, $NO\nu A$ and DUNE experiments, respectively. The CP violating parameter $\delta = 0$ and the matter density parameter $A \approx 1.01 \times 10^{-13}$ eV.  The solid (blue), dashed (red) and dot-dashed (black) correspond to the cases with $\delta =0$, $45^o$, and $90^o$, respectively. The maximum violation is indicated by the dashed horizontal.}
	\label{fig:W4}
\end{figure}

We now study the relevant Wigner and Clauser-Horne forms of LGIs for the case of neutrino system, keeping in mind the experimental constraints.  The inequalities should be casted in a form which is verifiable experimentally and at the same time leads to the maximum possible violation for the allowed  parameter range. It turns out that for the case of neutrino system Wigner form of LGI given by Ineq. (\ref{w1}) for the values of $m_{1}=-1$, $m_{2}= m_{3}=+1$ is most suitable. With initial neutrino state $\ket{\nu_\mu}$, we choose the dichotomic operator $\hat{A} = 2 | \nu_e \rangle \langle \nu_e | - \boldsymbol{I}$, where $\boldsymbol{I} = \sum_{\alpha=e,\mu,\tau}  | \nu_\alpha \rangle \langle \nu_\alpha |$. The operator $\hat{A}$ amounts to asking whether the neutrino is found in flavor $\nu_e$ ($+1$) or not ($-1$). With this setting, the standard LGI for three time measurement, turns out to be \( K_3 = 1 - 4\mathcal{P}_{\mu e}(t) + 4 \mathcal{P}_{e e}(t) \mathcal{P}_{\mu e}(2t) + 4 \beta(t) \), where $\beta(t)$ is a non-measurable term \cite{LGtIprd}. It is worth  noting here that for subatomic systems less number of measurements are preferable due to experimental constraints. Therefore, three time LGI is most relevant for such systems. In contrast to the standard LGI, one of the variants of Wigner form of LGI (denoted here by $W_Q$) turns out to be independent of non-measurable terms and can be shown to be

\begin{equation}
\label{W4}
W_{Q}=\mathcal{P}_{ee}(t) \mathcal{P}_{\mu e}(t) - \mathcal{P}_{\mu e} (2t) \le 0.
\end{equation}
Here, $\mathcal{P}_{\alpha \beta} (t)$ is the probability of transition from flavor state $\nu_\beta$ to $\nu_\alpha$ at time $t$. This is a remarkable coincidence which has the potential to have positive impact on experimental investigations in the context of LGI violations in neutrino oscillations. The behavior of $W_{Q}$  defined above is shown in Fig. (\ref{fig:W4}), in the T2K, NO$\nu$A, and DUNE setups with appropriate baseline and energy range.   The violation is more for DUNE experiment followed by NO$\nu$A and T2K, indicating that the long base and high energy experiments are more suitable for the experimental verification of these results.
\begin{figure}[t!]
	\centering
	\includegraphics[width=50mm]{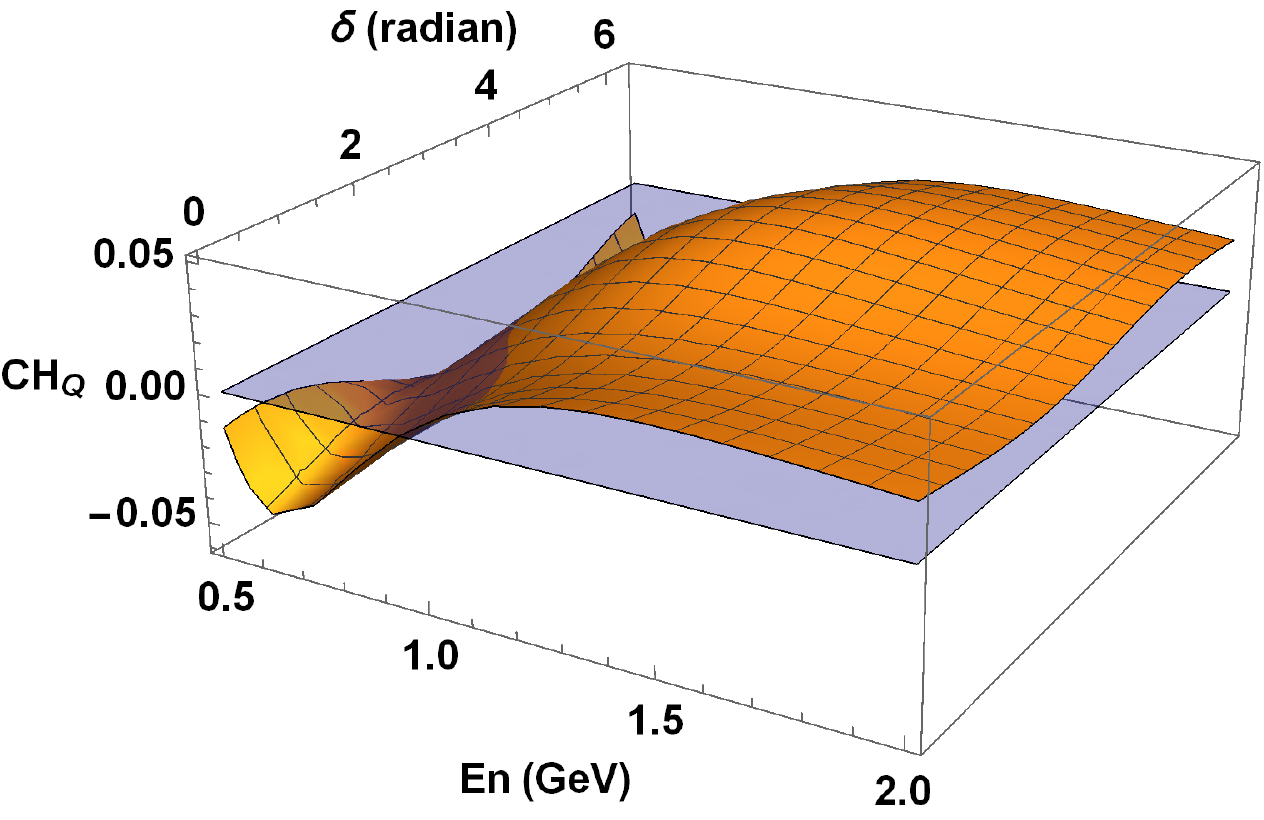}
	\includegraphics[width=50mm]{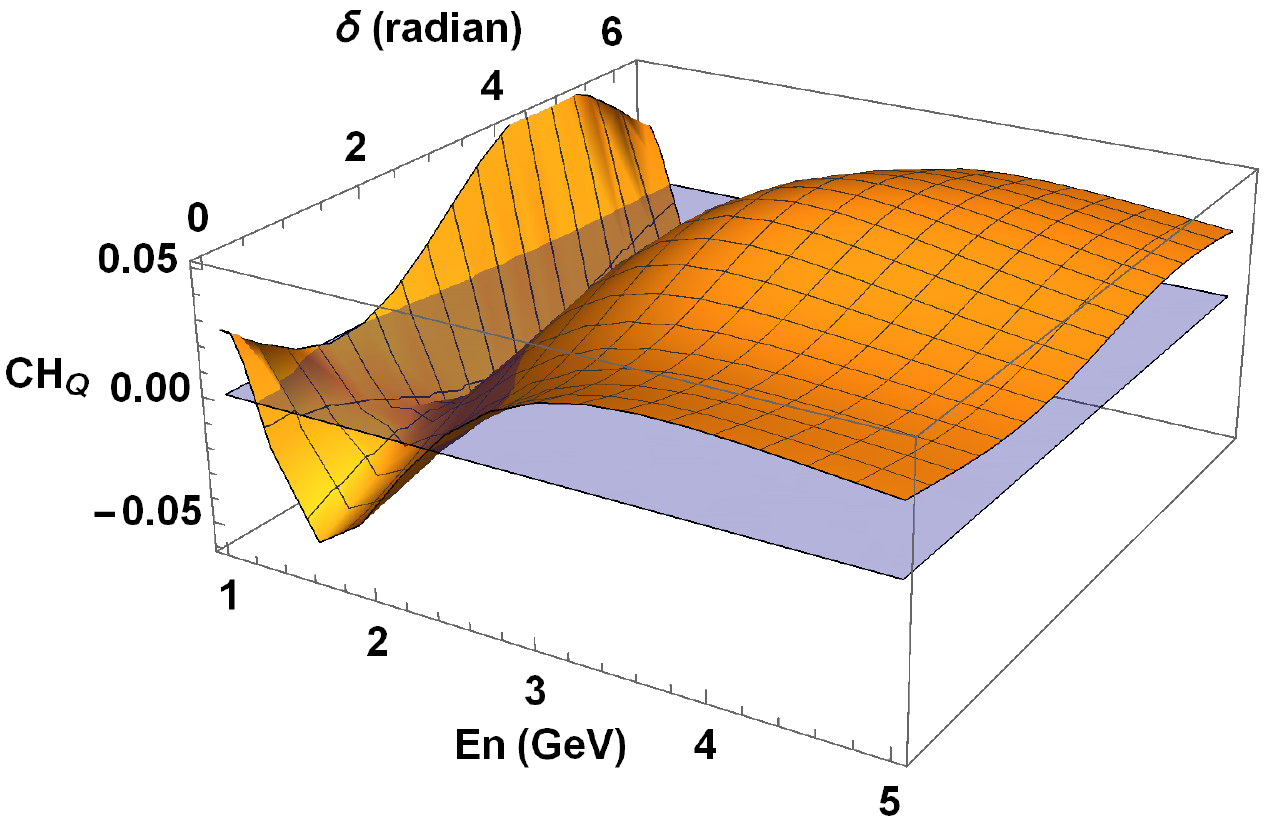}
	\includegraphics[width=50mm]{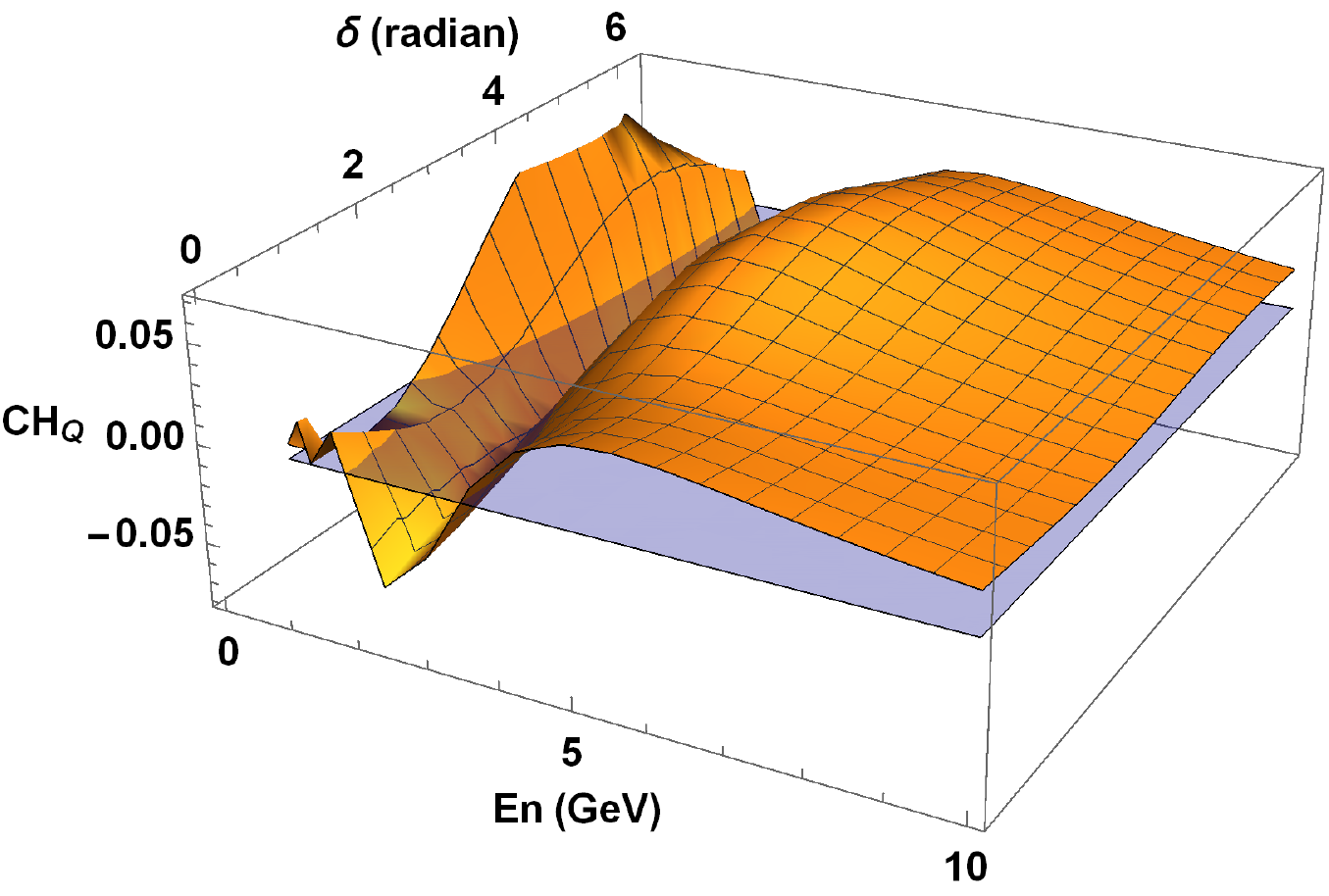}
	\caption{(color online) Clauser-Horne form of Legget-Garg inequality, Ineq. (\ref{ch4}), in neutrino system for different experimental set ups vz., $T2K$ (left), $NO\nu A$ (middle) and DUNE (right). The quantity $CH_{Q}$  is plotted with respect to the neutrino energy $E_n$ and the CP violating phase $\delta$.}
	\label{CHLGI4_Neutrino}
\end{figure}

\begin{figure}[t!]
	\centering
	\includegraphics[width=50mm]{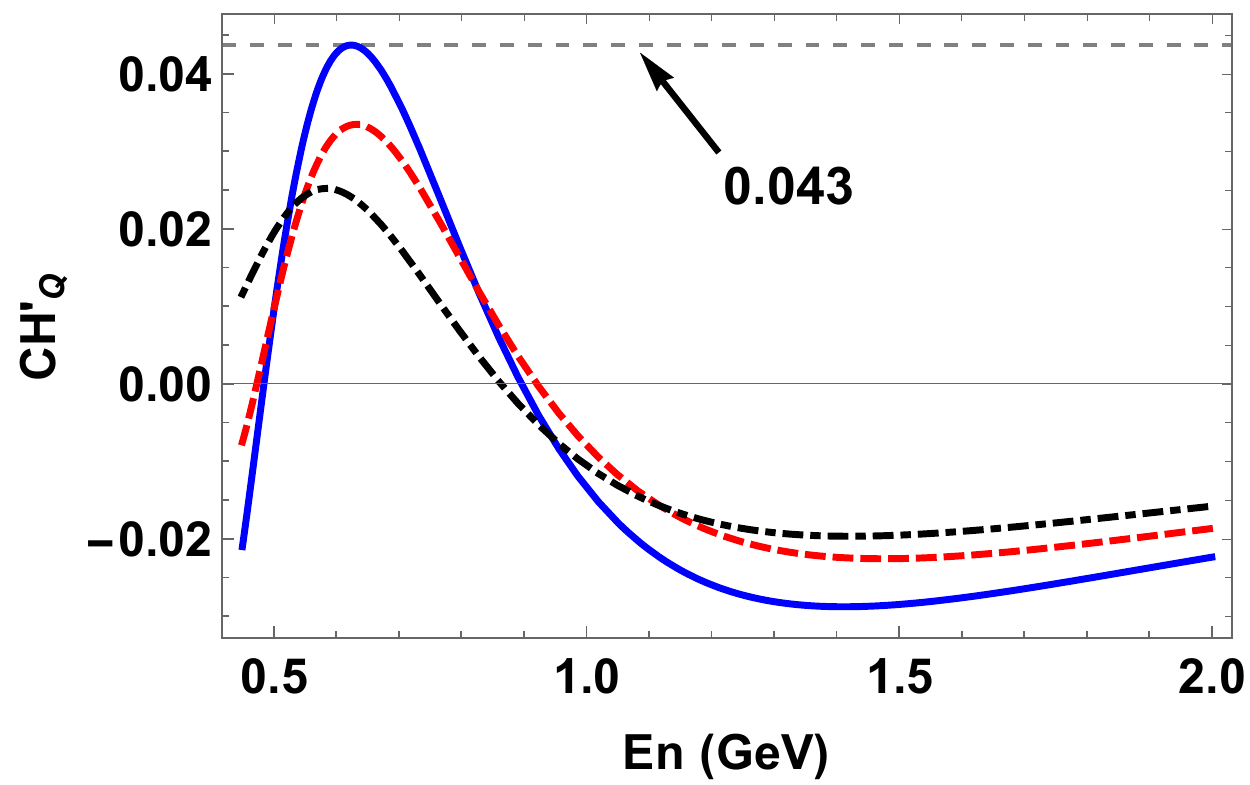}
	\includegraphics[width=50mm]{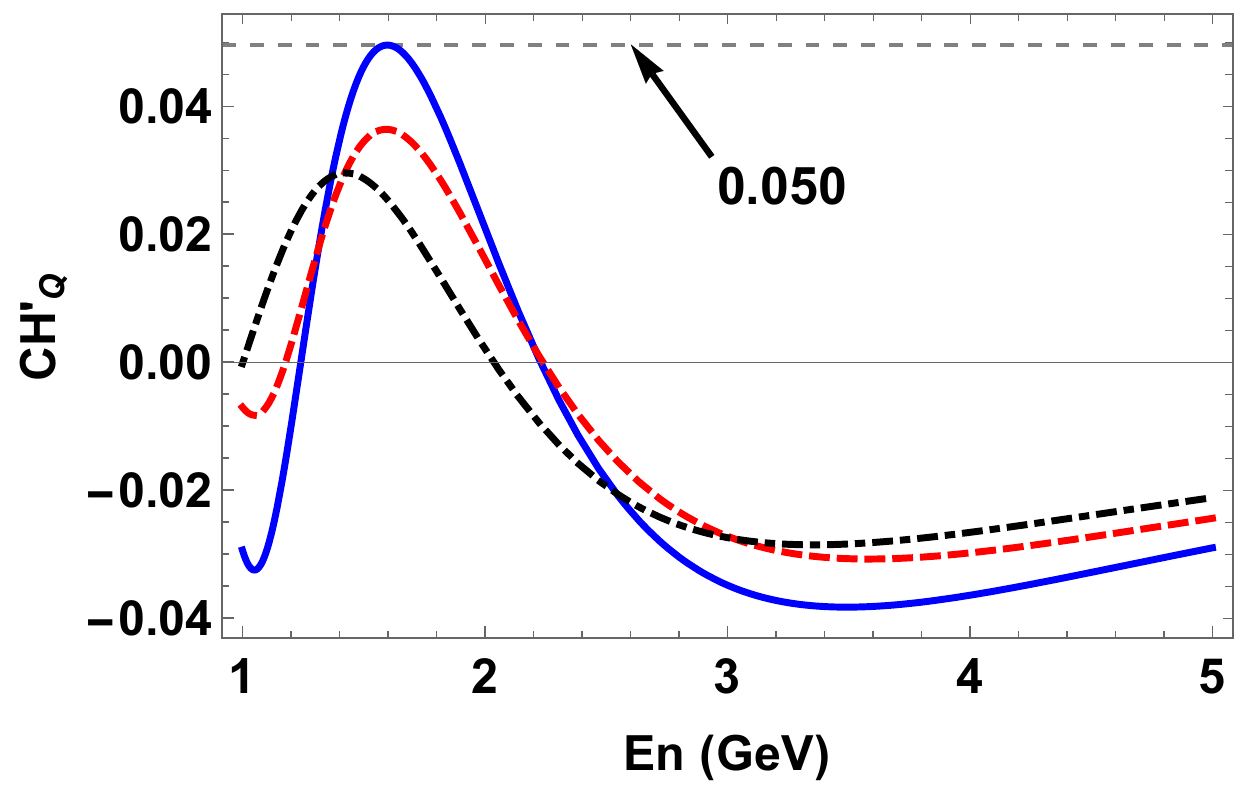}
	\includegraphics[width=50mm]{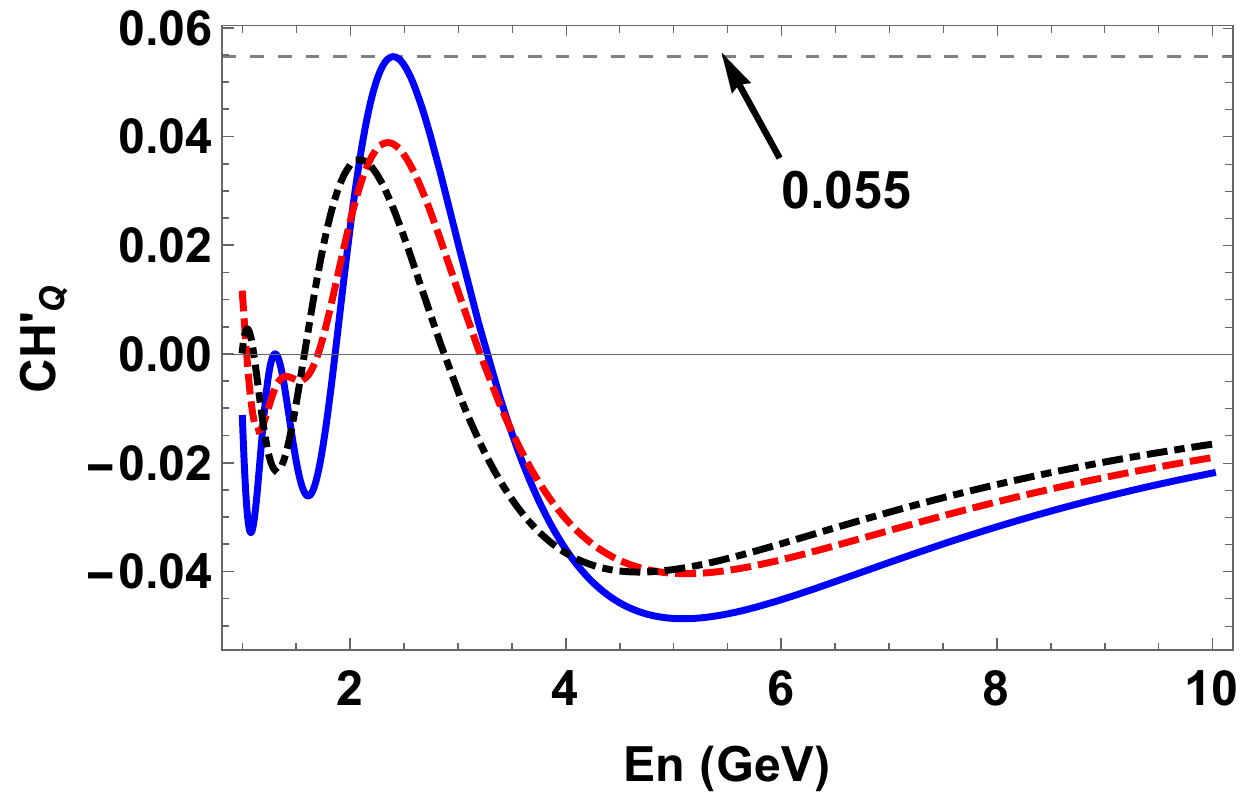}
	\caption{(color online) Clauser-Horne form of LGI ($CH^{\prime}_{Q}$), Ineq. (\ref{ch22}),  is depicted with respect to the neutrino energy $E_n$ in $T2K$ (left), $NO\nu A$ (middle) and DUNE (right) setups.  The presence of term $\mathcal{P}_{\tau \mu}$ makes the experimental verification of this quantity difficult in contrast  to the scenario depicted by Ineq. (\ref{ch4}).   The solid (blue), dashed (red) and dot-dashed (black) correspond to the cases with $\delta =0$, $45^o$, and $90^o$, respectively. The maximum violation is indicated by the dashed horizontal. }
	\label{CHLGI22_Neutrino}
\end{figure}
The suitable  Clauser-Horne form of LGI, can be found from the Ineq. (\ref{ch1}) for the values of $m_{1}=+1$, $m_{2}= m_{3}=-1$ and is denoted by $CH_Q$
\begin{equation}
\label{ch4}
CH_{Q} = -\mathcal{P}_{\mu e}(t)+\mathcal{P}_{ee}(t)\mathcal{P}_{\mu e} (2t) \le 0.
\end{equation}
Another useful Clauser-Horne form of LGI, $CH_Q^\prime$, can be obtained from the Ineq. (\ref{ch3}) for the values of $m_{1}=m_{3}=-1$, $m_{2}=+1$
\begin{align}
\label{ch22}
CH^\prime_{Q}&=\mathcal{P}_{\mu e}(t)- \mathcal{P}_{\mu e} (2t) [\mathcal{P}_{\mu e}(t)+\mathcal{P}_{\tau \mu}(t)] +\mathcal{P}_{\mu \mu}(2t)
+\mathcal{P}_{\tau \mu}(2t)-1\le 0.
\end{align}
The expressions for various probabilities appearing in the above equations can be seen from \cite{LGtIprd,naikoo2018entropic}. Figures (\ref{CHLGI4_Neutrino}) and (\ref{CHLGI22_Neutrino}) depict the behavior of $CH_{Q}$ and $CH^\prime_{Q}$, respectively, again for T2K, NO$\nu$A, and DUNE.  Here, it is important to note that the quantum violation of the Clauser-Horne form of LGI given by Ineq. (\ref{ch22}) is larger than the violation shown by the Ineq. (\ref{ch4}) and the Wigner form of LGI (Ineq.  (\ref{W4})) for the experimental set-up of DUNE. It is worth mentioning that $\mathcal{P}_{\alpha \beta} (t)$ depend, apart from time, on parameters like mixing angles, mass square difference, energy of the neutrino and CP violating phase (for $\alpha \ne \beta$).  In the ultra-relativistic limit,  time can be approximated by the distance it travels, i.e., $t \approx L$. Therefore, the Wigner parameter $W_Q$ becomes a function of $L$ and $2L$. This implies that an experimental verification of this inequality would require two detectors to be placed at $L$ and $2L$, respectively. However, in the present day experimental setups, such a provision is not possible. This difficulty can be bypassed by replacing the $2L$ dependence by $L$ in such a way that $\mathcal{P}_{\mu e} (2L, E)  =  \mathcal{P}_{\mu e} (L, \tilde{E})$ for energy $\tilde{E}$ within the experimentally allowed range.    Such an approach has been used  to study Leggett Garg inequality in the context of experimental facilities like  NO$\nu$A, T2K and DUNE \cite{LGtIneutrino}. It should be noted that for  vacuum oscillations, energies $E$ and $\tilde{E}$ are related by $\tilde{E} = E/2$. However,  this relation is not retained in the presence of matter effects.  Given that the matter modified oscillation probability is a smooth function of energy, it is always possible to find at least one $\tilde{E}$ which satisfies the above relation.  More explicitly, the solution of $\mathcal{P}_{\mu e} (2L, E)  =  \mathcal{P}_{\mu e} (L, \tilde{E})$ is obtained for a given value of the $CP$ violating phase within the energy window of the experimental setup. This obviously requires enough neutrino flux to make $\tilde{E}$ fall within the experimental regime. The DUNE experiment which has higher energy range is best suited for this approach.\par
In contrast to the standard LGI, an attractive feature of the  Wigner and Clauser-Horne forms of LGI is that some of these inequalities can be expressed completely in terms of measurable probabilities, as seen in Ineqs.  (\ref{W4}), (\ref{ch4}), and (\ref{ch22}), without invoking the stationarity assumption. However, Ineq. (\ref{ch22}) involves transition probabilities from flavor $\nu_\mu$ to $\nu_\tau$, which are beyond the scope of present experimental capabilities. The Wigner and Clauser-Horne forms of LGI may be  advantageous over the standard LGI, since the maximum violation occurs at energies around the maximum neutrino flux.   Further, Clauser-Horne forms show more violation in comparison to Wigner forms in the respective experiments as indicated explicitly in Figs. (\ref{fig:W4}) and (\ref{CHLGI22_Neutrino}).
\begin{figure}
	\centering
	\includegraphics[width=60mm]{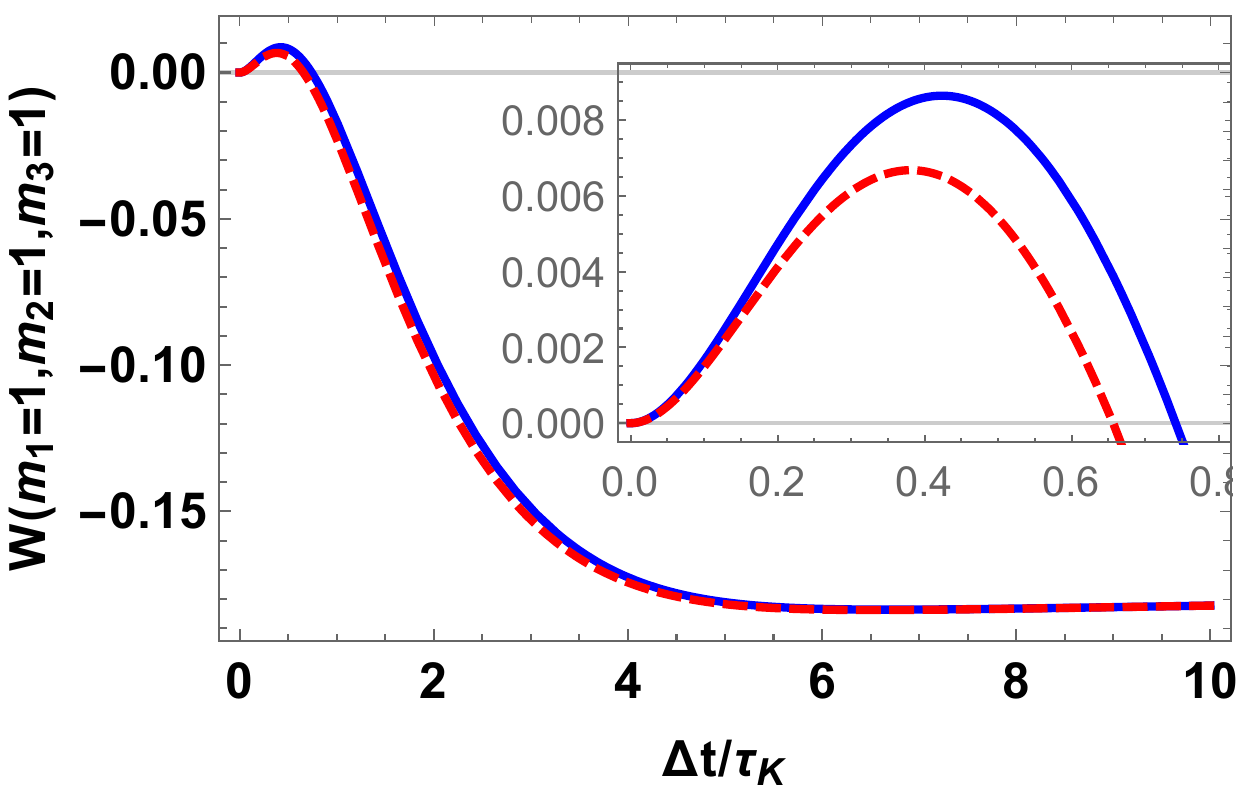}
	\includegraphics[width=60mm]{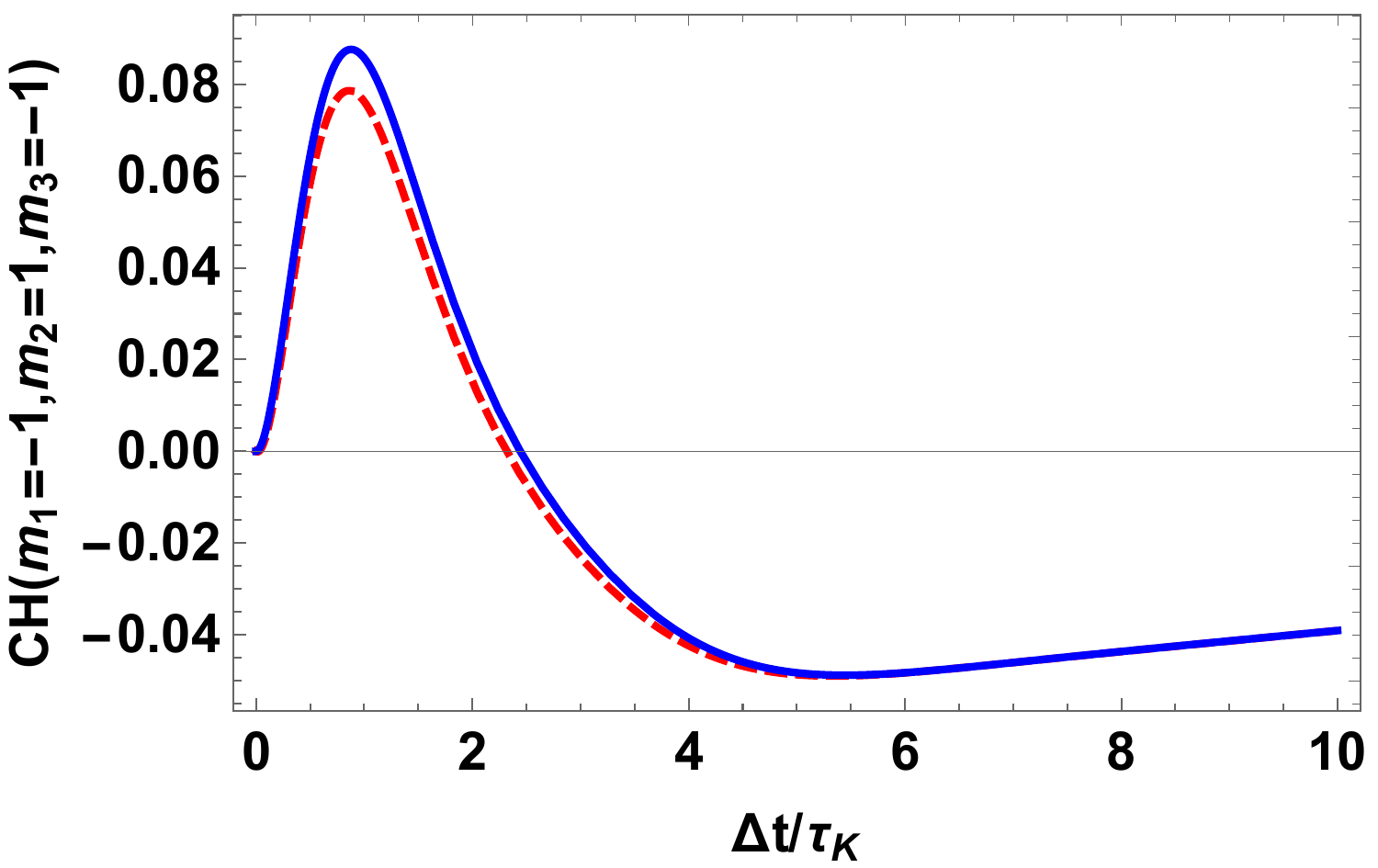}
	\caption{(color online) Wigner form of LGI (left), as given in Eq. (\ref{w1}) with  $m_{1}=m_{2}=m_{3}=+1$, and Clauser-Horne form of LGI (right), given by Eq. (\ref{ch1}) for $m_{1}=m_{3}=-1$ and $m_{2}=+1$,  are plotted with respect to the dimensionless parameter $\Delta t/\uptau_K$, where $\uptau_K$ is the average lifetime of $K^o$ meson and $\Delta t$ is the time interval between two successive measurements. Solid (blue) and dashed (red) curve corresponds to the case without and with decoherence, respectively. The effect of the decoherence is found to decrease the extent of violation, as expected.}
	\label{WLGI_CHLGI_Kmeson}
\end{figure}
\section{Quantum violation of Wigner and Clauser-Horne form of LGI in K-meson system}
Now, we discuss the relevant Wigner and Clauser-Horne forms of LGIs for the case of meson system which is inherently decaying in nature. The decoherence is controlled by the parameter $\lambda$ appearing in the Kraus operators. We assume that the initial state is $\ket{K^o}$ and the dichotomic operator is given by $\hat{O} = 2|K^o\rangle \langle K^o | - \mathbf{I}$, with $|K^o\rangle \langle K^o |+|\bar{K}^o\rangle \langle \bar{K}^o | + |0\rangle \langle 0 | = \mathbf{I}$. The operator $\hat{O}$ is $+1$  or $-1$ depending or whether the measurement outcome is $\ket{K^o}$ or not. After analyzing all the possible forms of  Wigner LGIs and Clauser-Horne form of LGIs for the meson system, we find the most appropriate is the one  given by Ineq. (\ref{w1}) for the values  $m_{1}=m_{2}=m_{3}=+1$. Further, the most suitable form of Clauser-Horne form of LGI for this system is given by the Ineq. (\ref{ch1}) for the values for $m_{1}=m_{3}=-1$ and $m_{2}=+1$. Unfortunately, the expressions for these inequalities turn out to be complicated, and are therefore not given here and are depicted numerically in  Fig. (\ref{WLGI_CHLGI_Kmeson}). However, it is worth pointing out here that the relevant expressions contain non-measurable terms. This can be surmounted by appealing to the LG type inequalities \cite{huelga1995proposed,huelga1996temporal} where the noninvasive measurability is replaced by \textit{stationarity} condition which is supported by the meson dynamics and has been studied in \cite{MesonPRD}.  As found in the case of neutrino system, the enhanced violations of Clauser-Horne form than Wigner form is again witnessed here with the former showing violations of around 10 orders of magnitude more than the later. Further, the effect of decoherence is expectedly reducing the extent of violation of the two inequalities. The various parameter (defined in Sec. (\ref{dynamics})) used in Fig.(\ref{WLGI_CHLGI_Kmeson}) are as follows: $\uptau = 1.889 \times 10^{-10}~s$, $\Gamma = 5.59 \times 10^9~s^{-1}$, $\Delta \Gamma = 1.1174 \time 10^{10}~s^{-1}$, $\lambda = 2.0 \times 10^{8}~s^{-1}$ and $\Delta m = 5.320 \times 10^{9}~s^{-1}$. Also, $Re[\epsilon] = 1.596 \times 10^{-3}$ and $|\epsilon| = 2.228 \times 10^{-3}$ \cite{d2006determination}.
\section{Conclusion}

 Given the interest in probing foundational issues in subatomic systems as well as the inherent difficulty in expressing the standard LGIs completely in terms of experimentally measurable quantities, in this work we study variants of Wigner and CHSH inequalities.
Neutrino dynamics is considered in  three flavor scenario including matter and CP violation effects. The meson system is treated using the open system formalism. For neutrino system,  it is found that  some of the  Wigner and Clauser-Horne forms of LGI  are more suitable in comparison to the standard LGIs from the experimental point of view, since these inequalities are in terms of experimentally measurable probabilities and the maximum violation is found to occur around the energies corresponding to the maximum neutrino flux. This  feature  should help in probing  foundational  issues in subatomic physics.  Specifically, we studied the violation of these inequalities in current running experiments like NO$\nu$A and T2K and also for the future upcoming experiment DUNE.  It turns out that the long base line and high energy experiments are more suitable for an experimental verification of such inequalities. Further in the context of mesons, treated using the open system formalism, the stationarity assisted LGIs is seen to be more suitable from the experimental point of view.  

In both neutrino as well as meson system, enhanced violation is found in the case of Clauser-Horne form of LGI as compared to Wigner form of LGI. Since, the extent of violation of various forms of LGI corresponds to the degree of quantumness of the system, therefore, decoherence is expected to reduce the extent of violation of these inequalities. These features are nicely manifested in the meson system. The optimal forms of various LGIs for either neutrinos or mesons are seen to depend on measurement settings.  This brings out the advantage of choosing appropriate LGIs and, therefore, provides scope for choosing various experimental setups for probing into foundational issues in subatomic physics.

\end{document}